**Practicality of Agent-Based Modeling of Civil Violence: an Assessment**

Christopher Thron (corresponding author), Elizabeth Jackson

Texas A&M University-Central Texas, Department of Mathematics

1001 Leadership Place, Killeen, TX 76549

thron@tamuct.edu

# 1 ABSTRACT:

Joshua Epstein (2002) proposed a simple agent-based model to describe the formation and evolution of spontaneous outbursts of civil violence (such as riots or violent demonstrations). Epstein's model uses two types of agents ("populace" and "cops"); and each populace agent is either active or quiescent according to its level of "grievance", its aversion to risk and the relative proportion of active agents and cops within a certain neighborhood of that agent. In this paper, we go a step beyond Epstein's study, and consider the application of the model to practical situations. First, we provide an improved mathematical characterization of the model dynamics that reduces the number of free parameters in the model. We show by simulations that there are no "tipping points" and the model behavior changes smoothly as a function of parameters. Using fundamental mathematical properties we explain the size and frequency distributions of outbursts, and confirm our results with simulation. We also argue that the model does not accurately reflect the mechanisms that lead to riot formation, and that nominal model parameters (such as "hardship", "legitimacy", and "average jail term") are proxies for complicated social processes; hence we conclude that the model is better suited as a phenomenological description rather than a theoretical explanation of crowd violence. Finally, we present a method for fitting the model to practical scenarios based on the frequency and severity of outbursts.

# 2 INTRODUCTION

Outbursts of civil violence (such as riots and violent demonstrations) occur worldwide due to a variety of causes that include political, social, ideological, and economic factors. Regardless of cause, such violence results in property damage and loss of life. One way to address this problem would be to develop a mathematical model which, based on currently available data, could forecast likely scenarios for the development of future outbursts. Such a model would be a valuable tool in the prevention and control of civil violence. This paper assesses the prospects of constructing such a model based on an agent-based model of civil violence outbursts originally proposed by Joshua Epstein [1].

The paper is organized as follows. Section 3 describes Epstein's model. Section 4 discusses mathematical properties of the model, and relates them to general features of model behavior. Section 5 shows the results of computer simulations using the model. Section 6 discusses prospects for using the model to simulate a practical system, and in particular examines issues relating to the determination of model parameters from practical socioeconomic data. Section 7 summarizes our conclusions.

## 3   BASIC MODEL

In Epstein's agent-based model [1], a region's population is represented as a set of agents located on a two-dimensional grid. Agents are of two types: *cop agents* and *populace agents*, where *populace agents* are further subclassified as belonging to different *groups* (which may reflect ethnicity or some other socio-economic grouping). At most one agent is located at every grid point. The grid is not fully populated (the maximum occupancy rate is typically less than 80 percent); empty sites are necessary to enable agent motion. The grid is initially populated so as to reflect the geographical distribution of the population groups being modeled.

The time dynamics of the model is specified for populace agents and cop agents in the following subsections.

### 3.1   Populace agent dynamics

At each time step, each populace agent is in one of three states: *active*, *inactive*, or *jailed*. Each populace agent that is jailed either remains in jail, or is released if its jail term comes to an end. Each populace agent that is not jailed makes an individual decision every time step whether or not to be active at that time step. This decision is based on the agent's perceived *grievance* ($G$) and *net risk* ($N$) according to the following criterion:

(1)  $G - N > T$,

where $T$ is a threshold that is the same for all agents for all time steps. The variable $G$ is supposed to reflect the agent's overall dissatisfaction level; while $N$ is intended to reflect the agent's disinclination towards activism due to fear of arrest.

The quantities $G$ and $N$ may be further broken down. In Epstein's model, the grievance $G$ depends on *hardship* ($H$) and *legitimacy* ($L$) according to the following relation:

(2)  $G = H \cdot (1 - L)$.

The variable $H$ is intended to reflect the degree of adverseness of the populace's socioeconomic condition, while $L$ reflects popular opinion of the government (so that $1 - L$ indicates the degree to which the populace blames the government for experienced hardship). Both $H$ and $L$ may vary from individual to individual: in the model, this is reflected by choosing $H$ and $L$ randomly according to fixed probability distributions. Ordinarily, both $H$ and $L$ are restricted to the interval [0,1], although hardship need not be restricted to this interval. In our implementation both $H$ and $L$ are chosen randomly for each agent from a uniform distribution on fixed intervals: for $H$, the interval varies from group to group, while for $L$, the interval is fixed for the entire population.

The net risk $N$ depends on the agent's estimated arrest probability $P$ and risk aversion $R$ according to the formula

(3)  $N = P \cdot R$.

Each agent's estimated arrest probability is computed based on the distribution of cops and active agents within a certain (Euclidean) distance of that agent: this distance is a model parameter called the populace agent *vision* ($V_a$). The agent computes the ratio of number of cops within vision to number of active agents within vision (where for purposes of computing the ratio, the agent itself is counted as active): this ratio is denoted as $(C/A)_v$. In Epstein's original paper, the populace agent's estimated probability of arrest is specified as

$$(4)\ P = 1 - (0.1)^{(C/A)_v},$$

which Epstein derives on the basis of a "rational" populace agent model. However, the most widely-distributed version of Epstein's model (posted on the NetLogo web site by U. Wilensky[2]) uses the function

$$(5)\ P = 1 - (0.1)^{\lfloor (C/A)_v \rfloor},$$

where $\lfloor \ldots \rfloor$ denotes the "floor" function. Wilensky justifies this function empirically rather than theoretically; he observes that unlike (4) it produces outbursts of activism. In Section 4.1 we propose alternatives that have both theoretical and empirical support.

The per-agent parameters are summarized in Table 1.

Table 1 Populace agent parameters

| Parameter | Denotation | Determination method (for each populace agent) |
|---|---|---|
| G | Grievance | Equation: $G = H(1-L)$ |
| H | Hardship | Uniform distribution on $[H_0 - \Delta H, H_0 + \Delta H] \subset [0,1]$ (interval may vary from group to group) |
| L | Legitimacy | Uniform distribution on $[L_0 - \Delta L, L_0 + \Delta L] \subset [0,1]$ |
| N | Net risk | Equation: $N = RP$ |
| R | Risk avoidance | Uniform distribution on $[0,1]$ |
| P | Estimated arrest probability | Equation (depends on $(C/A)_v$) |
| $V_a$ | Agent vision | Fixed value, same for all agents |

The agents also move at each time step. The motion is specified so that the agent's new position can be any of the points within $V_a$ of the agent's current position, chosen with equal probability.

### 3.2 Cop agent dynamics

Besides populace agents, the model also includes "cop" agents, that "arrest" active agents they come across. A *cop vision* parameter $V_C$ is defined, and $V_C$ is the same for all cop agents. In his original paper, Epstein consistently sets $V_c = V_a$, and in our study we follow this same practice.

For each cop agent each time step, if any of the populace agents within a distance $V_C$ of the cop agent is active, then the cop agent will choose one of these active populace agents with equal probability. In this case, the cop agent moves to the location of the chosen active populace agent, which then changes its status to "jailed" and is assigned a jail term (chosen uniformly randomly from 0 to a maximum value $J_{max}$). On the other hand, if no populace agents within a distance $V_C$ of the cop agent are active, then the cop agent moves to one of the points within $V_c$ of its current position, chosen with equal probability.

We note that this specification of cop agent dynamics is slightly different from that in the NetLogo implementation, which has each cop agent move to another point within $V_c$ before looking to arrest an active agent within Vc of this new point. We considered the NetLogo implementation to be unrealistic since the cop agent essentially moves twice within each time step, doubling its mobility.

In our implementation the following restriction was imposed on cop agent motion: if a cop agent does not arrest an active agent, then it is required to move to an empty location that is *next to* a populace agent.

This change prevents cops from wandering away from the population, which can easily happen if the populace agents are not spread uniformly over the grid. When the grid is uniformly populated, the effect of this modification is greatly lessened.

## 4 MATHEMATICAL CONSIDERATIONS

### 4.1 Model characterization

In this section, we develop a simpler characterization of the agent-based model which enables us to better identify and classify model behavior.

According to equations (1)–(3), several factors go into each agent's decision on whether or not to be active, including the agent's values of hardship, legitimacy, risk aversion, and arrest probability ($H, L, R,$ and $P$ respectively). $H,L,R$ represent properties internal to the agent; and since populace agents are placed randomly, these internal properties are randomized over space-time locations.[1] $P$ depends on the current neighboring conditions, namely the value of $(C/A)_v$ (or equivalently, $(A/C)_v$) It follows that the populace's statistical behavor is completely determined by *activist response function f*, defined as follows:

$f(y)$ = Probability that a randomly-generated populace agent will choose to go active under the condition that $(A/C)_v = y$

The function $f(y)$ corresponds to the fraction of populace agents that respond actively to the condition $(A/C)_v=y$, where populace agent response is averaged over all the joint probability distribution on $H, L,$ and $R$. In this definition we have used $(A/C)_v = y$ rather than $(C/A)_v = y$, since it turns out that the results are easier to interpret practically with this choice. Figure 1 shows examples of different possible $f(y)$ curves: all curves in the figure correspond to the same set of population parameters.

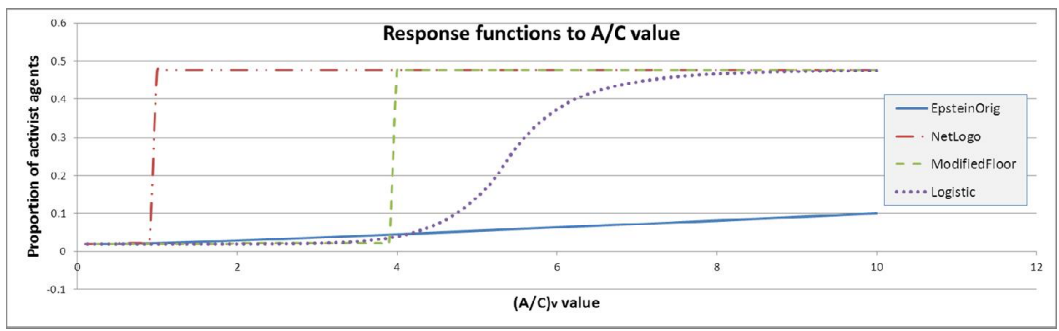

Figure 1 Activism response function curves. "EpsteinOrig" corresponds to the model described in [1]; "NetLogo" corresponds to the model implemented in [2]; "Modified floor" is similar to the NetLogo version, but with a different critical crowd size; and "Logistic" is the function implemented in this paper. All functions correspond to the same set of population parameters.

Note that all $f(y)$ curves in Figure 1 have the same minimum and maximum values, which correspond to $y \to 0$ and $y \to \infty$ limits respectively. These are related to model parameters as follows:

- $f(y) \to (G_{max} - T)^2 / (2G_{max})$    as  $y \to 0$,

---

1 Actually, in a mathematical sense the agents are not fully randomized; but the random placement of agents guarantees a close approximation to complete randomization, particularly for large systems.

- $f(y) \to 1 - T/G_{max}$                  as $y \to \infty$.

These limits have the following practical interpretations:

- A fraction $(G_{max} - T)^2 / (2G_{max})$ of populace agents is active unconditionally, regardless of local conditions. We may thus identify $(G_{max} - T)^2 / (2G_{max})$ as the fraction of populace agents that are *extremists*.
- $1 - T/G_{max}$ is the maximum fraction of the population that can ever become active. In other words, a fraction $T/G_{max}$ of populace agents is *inert*: these agents will never become active under any circumstances.

In summary, the response function $f(y)$ completely characterizes the statistical response of populace agents. In turn, $f(y)$ can be characterized by the following four variables:

- Extremist fraction: fraction of populace agents that is never inactive. The extremist fraction is equal to $f(0)$.
- Inert fraction: fraction of populace agents that are never active. The inert fraction is equal to $1 - \lim_{y \to \infty} f(y)$.
- Critical crowd size: activist/cop ratio above which $f(y)$ experiences a sharp increase. The critical crowd size is 1 for the NetLogo version, and is equal to $N$ for both modified floor and logistic versions. The original Epstein version has no critical crowd size (and hence exhibits no outbursts).
- Saturation crowd size: activist/cop ratio at which $f(y)$ reaches 95% of its maximum value. For the NetLogo and modified floor versions, the saturation crowd size is equal to the critical crowd size. For the logistic curve shown in Figure 1, the saturation crowd size is about 7. For a general logistic curve, the saturation crowd size is controlled by the exponent $p$ (larger values of $p$ will produce smaller saturation crowd sizes).

These four variables essentially characterize the shape of $f(y)$, which in turn determines the statistics of populace agent activist response. In Section 5 when we characterize model behavior's dependence on parameter values, we will investigate the system's response to three of these four variables (we do not include the saturation crowd size in our study, because it is closely related to the critical crowd size).

### 4.2 Mathematical properties and implications

In this section we discuss important general statistical properties of the model that follow directly from the fundamental mathematical structure of the model:

- First, the model is a *Markov Chain*. We will show that this implies that the probability distribution of the time between outbursts is approximately *exponential*, so that future outbursts are essentially unpredictable based on past outburst history, in a sense to be explained precisely below.
- Second the model exhibits *self-organized criticality (SOC).* As a result, outburst sizes for any particular model realization roughly follow a power-law distribution (similar to the distribution of earthquake sizes).

In the following discussion we will provide theoretical and empirical justification for these claims.

### 4.2.1 Markov property and consequences

A Markov chain is a system where the probability of various future states only depends on the present state of the system. As a result of the Markov property, it is possible to show [4] that past a certain time window the waiting time between outbursts of activism is *memoryless*. In other words, if sufficiently long time has elapsed since the last outburst, then the amount of time since the last outburst gives *no information* about the next outburst. We may compare this situation to repeated throwing of a dice or coin. The probability that the next flip of a coin will be heads is still ½, no matter how many previous consecutive coin-flips are tails: similarly, the probability that an outburst occurs at the next time step is a constant, independent of the number of time steps since the last outburst. This result underscores the unpredictable nature of the next outburst occurrence. It also implies that the probability distribution for the waiting time between outbursts can be characterized by a single number, namely the probability of outburst per step.

A simplified conceptual picture of the agent-based model may be given by dividing the states into three sets (this division is heuristic, and is intended only as a conceptual aid):

- $Q'$ denotes the set of states of the system that correspond to an ongoing outburst of activism (for instance, any state that includes a large number of activist agents will belong to $Q'$).
- $Q''$ denotes system states associated with recovery from an outburst (for instance, any state where a large number of agents are in jail).
- $Q$ denotes the remainder of the states: these can be considered as "normal" states of the system.

We may consider the evolution of the system which passes through an outburst of activism in terms of these three sets. During the outburst, the system's state will be in the set $Q'$. When the outburst subsides, the Markov chain usually passes into the set $Q''$; and following recovery, the system usually passes into the set $Q$, where it stays until the next outburst. We use $Q^*$ to denote the event that the system enters the next outburst after the first outburst has subsided (note the system configurations represented by $Q^*$ are identical to those in $Q'$: we use a different symbol to distinguish between outbursts). Besides the usual transitions, we also allow transition probabilities from $Q'$ to $Q^*$ and from $Q''$ to $Q^*$. This leads to the state transition diagram shown in Figure 2.

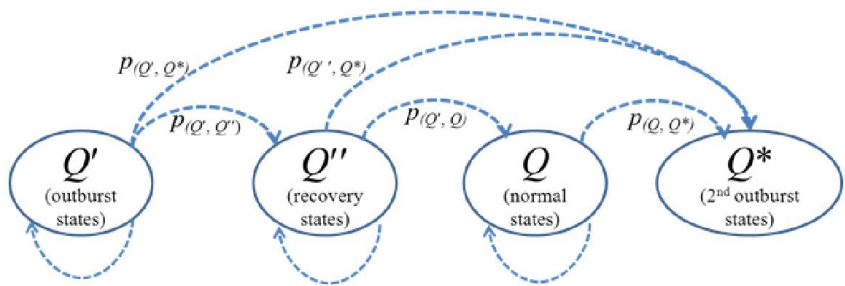

Figure 2 State diagram for modified simple Markov model

The resulting outburst waiting time from a modified simple Markov model is shown in Figure 3, together with Epstein's empirical results. The close qualitative agreement supports the validity of the transition picture given in Figure 2.

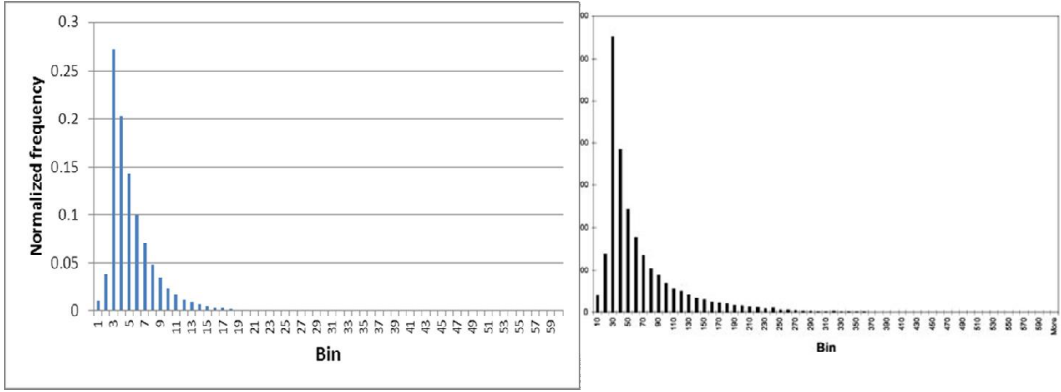

**Figure 3** (Left) Distribution of outburst waiting time from Markov model; (right) Outburst waiting times, from Epstein' paper [1].

### 4.2.2 Self-organized criticality and power-law size distributions

Self-organized criticality (SOC) as a common property of phenomena involving large ensembles of individuals was first identified by Bak, Tang, and Wiesenfeld [5]. In their original paper, they illustrated SOC using a *sandpile model*. One particular example of a sandpile model is the *abelian sandpile*[6], described as follows. Grains of sand are located at grid points on a $N \times N$ grid. The number of grains of sand piled at grid location $(i,j)$ is denoted as $n_{i,j}$. At each time step, a grain of sand is added to a randomly-chosen grid point $(i^*,j^*)$. If $n_{i^*,j^*}$ then exceeds a fixed threshold $K$, then a grain of sand from location $(i^*,j^*)$ is sent to each neighboring grid point. This model produces avalanche-like behavior, where the addition of a single grain of sand can affect a large area of the grid.

There is no rigorous mathematical definition of the SOC property. Instead, SOC systems are characterized by certain key features, including the following:

- Instability effects are instigated at random sites in the grid;
- Instability occurs suddenly (not gradual onset);
- Instability propagates rapidly to neighboring sites;
- Instability conditions at the original site are relaxed.

These features are illustrated schematically in Figure 4.

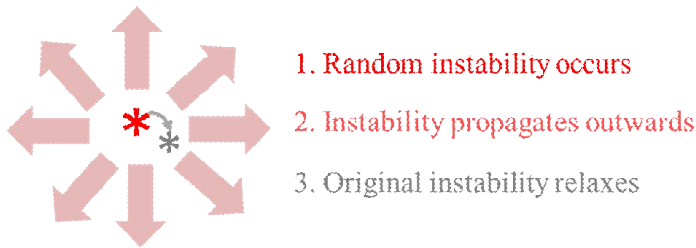

Figure 4 SOC model properties

All of these conditions hold for the agent-based Epstein model, if we replace individual grid points in the sandpile model with *vision circles*, where a vision circle consists of all sites within a (Euclidean) distance of one agent vision of a particular grid point. Figure 5 shows how the agent-based model exhibits all of these SOC features. First, instability occurs randomly due to fluctuations in the presence of cops: when cops happen to be absent from a region larger than a vision circle, then agents near the center of that region will have $(A/C)_V \to \infty$, which means the maximum fraction possible of those agents will become active. This creates a "knot" of active agents. If this knot exceeds the critical crowd size $N$ (described in Section 4.1 above), then other agents in neighboring vision circles become active even when cops are within their vision. This is one mechanism by which an outburst propagates to neighboring circles. Another mechanism is due to the fact that cops in neighboring vision circles move to arrest the original knot of active agents, reducing the density of cops in surrounding areas and promoting activism in neighboring circles. Finally, the incoming cops arrest the original activists, causing a lessening of the original outburst.

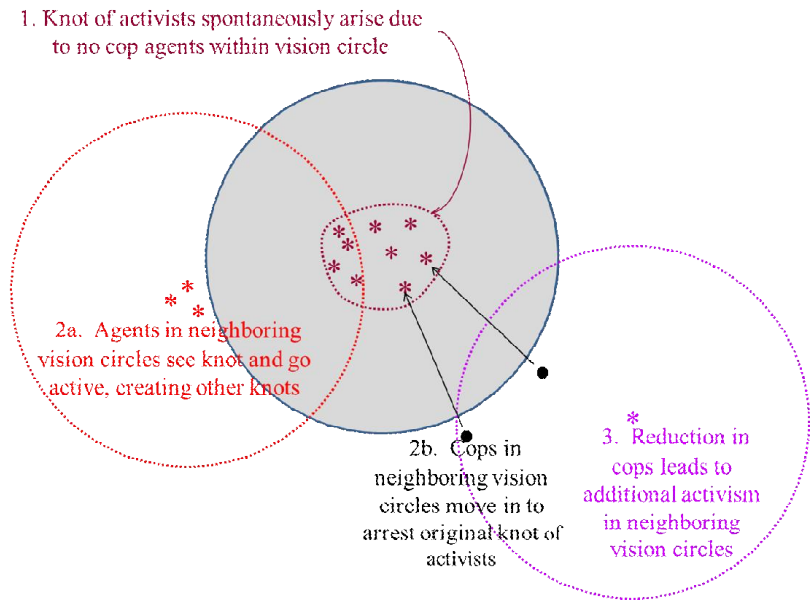

Figure 5 SOC properties of agent-based model

One of the key mathematical consequences of SOC is power-law size distributions: that is, the probability distribution of outbursts obeys the following distribution:

(6) $\Pr[\text{outburst size} = Z] \propto Z^{-a}$,

where $a$ is the *exponent* of the distribution (typically $\alpha > 1$). This relation can also be conveniently expressed as

(7) $\log \Pr[N \leq \text{outburst size} \leq 10N] \approx K - \beta \log N$      (power law with exponent $1+\beta$, $\beta > 0$)

or in the case where $\alpha = 1+\beta \approx 1$,

(8) $\log \Pr[N \leq \text{outburst size} \leq 10N] \approx K$,      (power law with exponent 1)

where $K$ is a constant that depends on $\alpha$. Power law distributions have "fat tails", in the sense that large events are more likely than in other common distributions. For example, compare (7) and (8) with the corresponding relation for an exponential distribution:

(9) $\log \Pr[N \leq \text{outburst size} \leq 10N] \approx K - N$      (exponential distribution)

We found support for power-law size distributions in practical data. Figure 6 displays riot data obtained for Nigeria and for all Africa from the ACLED dataset (www.acleddata.com). Both show much more robust tails than would be predicted by exponential distributions: this characteristic of power-law dependence.

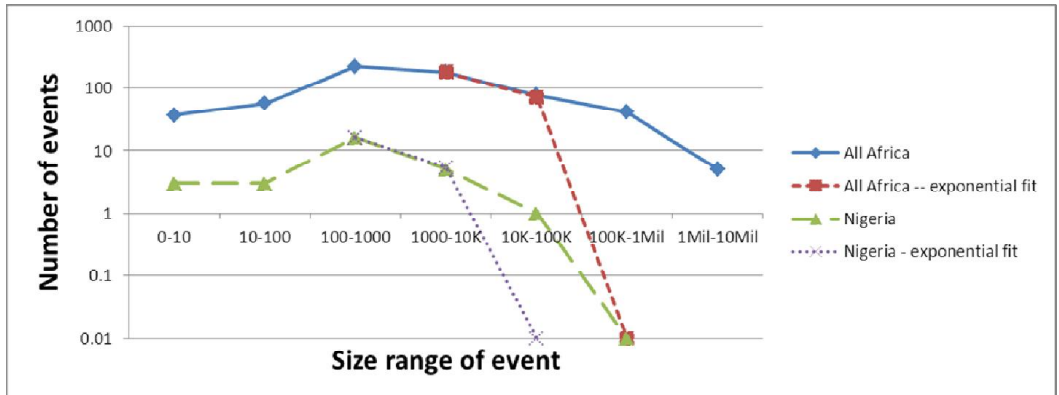

Figure 6 Number of riot events versus size range for riots in Africa and Nigeria (1997-2011), from ACLED dataset (www.acleddata.com)

## 5 SIMULATIONS AND PARAMETER MAPPING

### 5.1 General model behavior

When running the model on the computer, in order to represent the activity of the system over time we recorded the number of active agents at each time step, then plotted the results on a graph. Due to the randomness involved in the agents' dynamics, the system behavior was different for each simulation run. Qualitative and statistical features were similar from run to run as long as simulation parameters were not changed; but changes in the simulation parameters led to significant changes in the statistics.

Figure 7 shows levels of activism versus time four individual simulations at four different values for number of cop agents. At large cop densities (blue line) no outbursts occurred. As the number of cop agents was decreased outbursts begin to appear, which increase steadily both in size and frequency as the number of cop agents decreases further. At low cop densities, outbursts attain a maximum size, and there is much greater uniformity in the size of theses "superbursts" than at lower cop density levels. This behavior (which we denote as "hyperactivity") was also observed by Epstein. It appears that since his agent response curve is a step function rather than a smooth curve (see Figure 1), he did not observe the smooth transition leading up to hyperactivity.

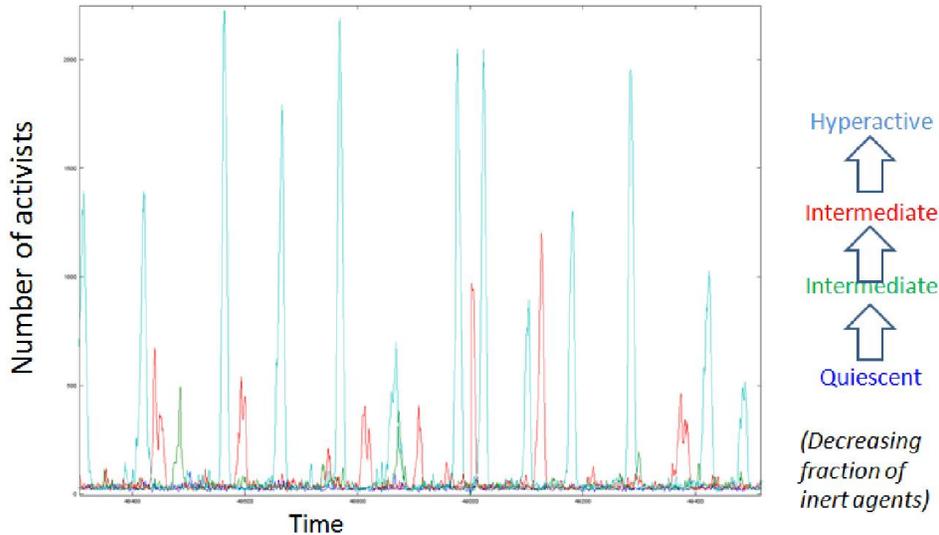

Figure 7 Active agents for systems at different cop agent densities. Behavior varies continuously from quiescent at high cop densities (blue) to hyperactive (teal) at low cop densities

## 5.2 Outburst dependence on excitable fraction critical crowd size, and extremist fraction,

In Section 4.1 we defined the population parameters *excitable fraction, critical crowd size*, and *extremist fraction.* In this subsection we describe the model's behavior as a function of these three parameters, as determined by simulations.

Table 2 gives the basic parameters that were common to all simulations. The excitable fraction, critical crowd size, and extremist fraction were then varied independently in different situations: the values used are shown in Table 3, and the corresponding response functions $f((A/C)_v)$ are shown in Figure 8, Figure 9 and Figure 10 for excitable fraction, critical crowd size, and extremist fraction respectively. In each group of simulations, only one of the three variable parameters was varied between simulations, while the other two variables were maintained at index #3 values.

**Table 2 Parameters common to all simulations**

| Simulation Parameter | Value |
|---|---|
| Grid size (cells per edge) *(grid is square)* | 80 |
| Number of populace agents | 4480 |
| Number of cop agents | 152 |
| Cop / agent vision | 7 |
| Maximum jail term | 100 |
| Simulation length | 1000 |
| Number of simulations per parameter set | 200 |

**Table 3 Parameter values used in simulations (varied separately)**

| Parameter | Inde | Inde | Inde | Inde | Inde |
|---|---|---|---|---|---|

| Excitable | 0.6 | 0.7 | 0.8 | 0.9 | 1.0 |
|---|---|---|---|---|---|
| Critical crowd | 4.2 | 4.1 | 4.0 | 3.9 | 3.8 |
| Extremist | 0.01 | 0.02 | 0.02 | 0.03 | 0.03 |

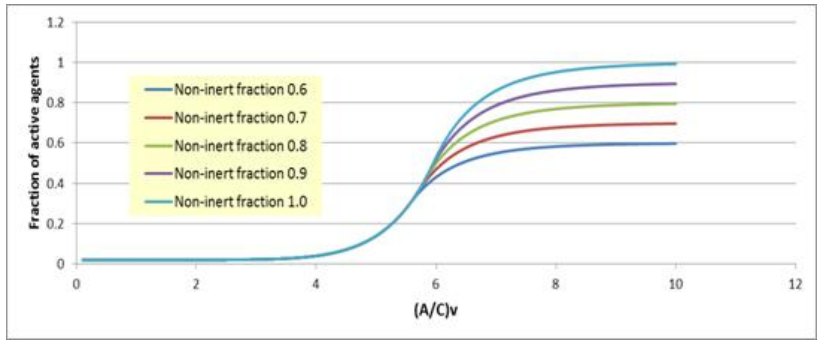

**Figure 8 Agent response functions for different excitable (non-inert) fractions used in simulations**

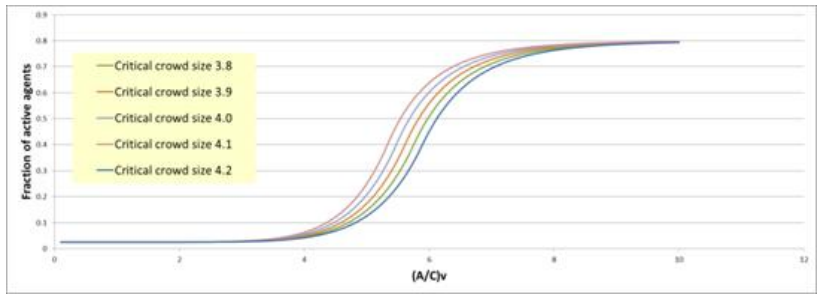

**Figure 9 Agent response functions for different critical crowd size levels used in simulations**

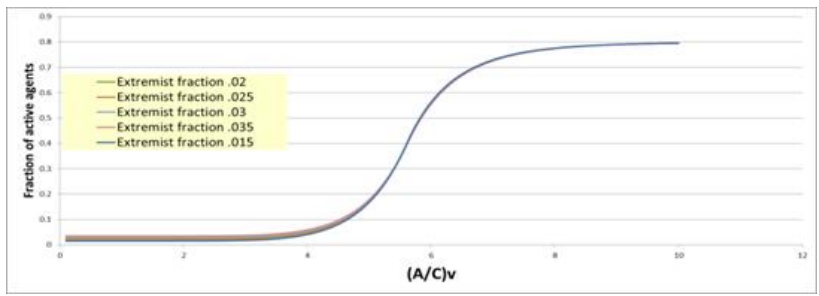

**Figure 10 Agent response functions for different extremist fractions used in simulations**

Figure 11 shows the base 10 logarithm of mean outburst frequency as a function of excitable fraction (left), critical crowd size (center), and extremist fraction (left). The different lines in each graph (labeled 64, 96, 144) correspond to different thresholds for minimum outburst size – for example, at the '64' level an outburst was logged as initiated whenever the number of active agents crossed from below 64 to above 64. The outburst was then logged as terminated when the number of active agents next dropped below 45.

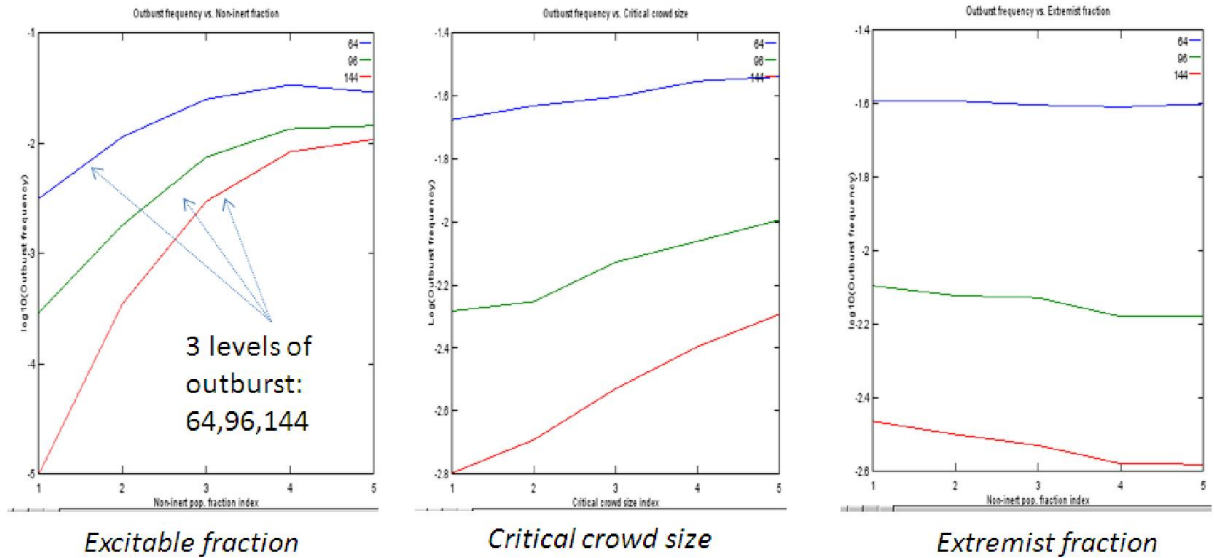

**Figure 11 Frequency of outbursts (log scale) for increasing values of excitable (non-inert) fraction (left), decreasing critical crowd size (center), and increasing extremist fraction (left). The three different curves in each graph correspond to different thresholds for minimum outburst size.**

The leftmost graph of Figure 11 shows that the excitable fraction had a huge effect on outburst frequency. Note that one tick mark on the vertical (log) scale corresponds to a 10× increase in frequency. This means that, for example increasing of excitable fraction from 0.6 to 1.0 caused a change in outburst frequency at the 144 level from $10^{-5}$ to $10^{-2}$, which is about a 1000× increase. Interestingly the curve is *concave*, indicating that increasing the excitable fraction had a *decreasing* relative effect on outburst frequency. In fact, excitable fractions of 0.9 and 1.0 gave nearly the same mean outburst frequencies, regardless of threshold. For a threshold of 64, the mean frequency is roughly $10^{-1.5}$, so the mean time between outbursts is about 30, which somewhat less than the mean jail term (50) for the simulations. It seems that for this set of parameters the situation is so volatile that successive outbursts begin even before jailed activists from the previous outburst have a chance to get out of jail (since the grid is large, the outbursts take place in different locations on the grid).

The middle graph of Figure 11 shows that decreasing the critical crowd size also significantly increased outburst frequency: a 10% decrease in critical crowd size led to a roughly 25% increase in outburst frequency, regardless of threshold.

The rightmost graph of Figure 11 gives an entirely unexpected result: increasing the fraction of extremists actually led to a slight *decrease* in the frequency of outbursts! The effect is slight: doubling the extremist fraction produced a decrease of roughly 5% in the frequency of outbursts. This paradoxical result may possibly be explained by the fact that extremists in the model are almost always in jail, since they protest even when they have no support from neighboring agents and are thus immediately re-arrested after they are released from jail. It is questionable whether such agents exist in practical situations: so it seems that this paradoxical behavior is due to unrealistic features of the model.

Figure 12 shows the base 10 logarithm of mean outburst peak size as a function of excitable fraction (left), and critical crowd size (center), and extremist fraction (right). The outburst peak size is defined as the maximum number of activists per time step occurring during an outburst. The different lines in each graph (labeled 64, 96, 144) correspond to different thresholds for minimum outburst size, just as in Figure 11.

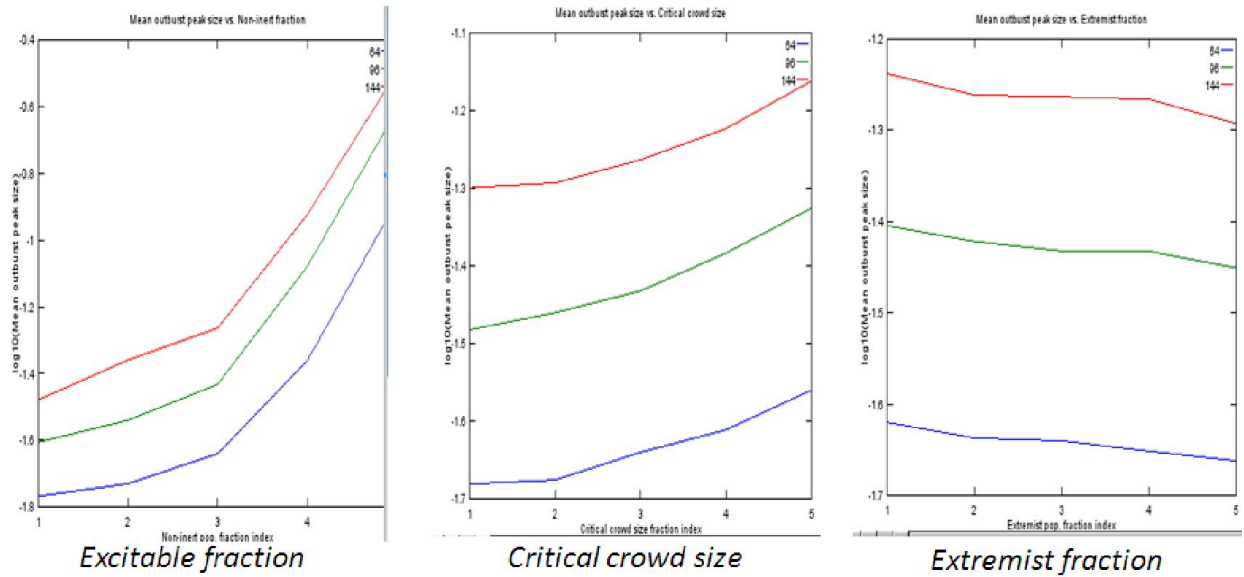

**Figure 12 Mean outburst peak size (as a fraction of the total population) for different levels of excitable fraction (left), critical crowd size (center) and extremist fraction (right). The different curves correspond to different thresholds for minimum outburst size.**

The leftmost graph of Figure 12 shows that in contrast to the outburst frequency the excitable fraction is *convex*, indicating that increasing the excitable fraction has an *increasing* effect on the logarithm of outburst peak size. This means that successive increases in the excitable population can cause disproportionately large increases in the size of protests. For example, a 12% increase in excitable fraction from 0.8 to 0.9 causes an increase of roughly 0.3 in the base 10 logarithm of outburst peak size, which translates to a $10^{0.3} \times 100 \approx 200\%$ increase in maximum protest participation! Although the graphs do not show a "tipping point" or specific set of conditions where protests begin to occur, it does show that past a certain point small increases in the excitable population lead to huge increases in protest size.

The middle graph of Figure 12 shows that the effect of a decrease in critical crowd size on peak outburst size is similar to its effect on outburst frequency: a 10% decrease in critical crowd size led to a roughly 25% increase in outburst peak size, regardless of threshold.

The rightmost graph of Figure 12 parallels the unexpected result shown in Figure 11: increasing the fraction of extremists from 0.035 to 0.015 slightly decreases the outburst peak size. As before, this may possibly have something to do with the fact that extremists are largely jail-bound, so are typically unable to participate in protests.

Figure 13 shows the base 10 logarithm of mean number of arrests as a function of excitable fraction (left), and critical crowd size (center), and extremist fraction (right). The number of arrests is counted as a fraction of the total population: so for instance a value of -0.5 on the vertical scale means that a fraction of $10^{-0.5} \approx 0.3$ of the total population is arrested per protest on average.

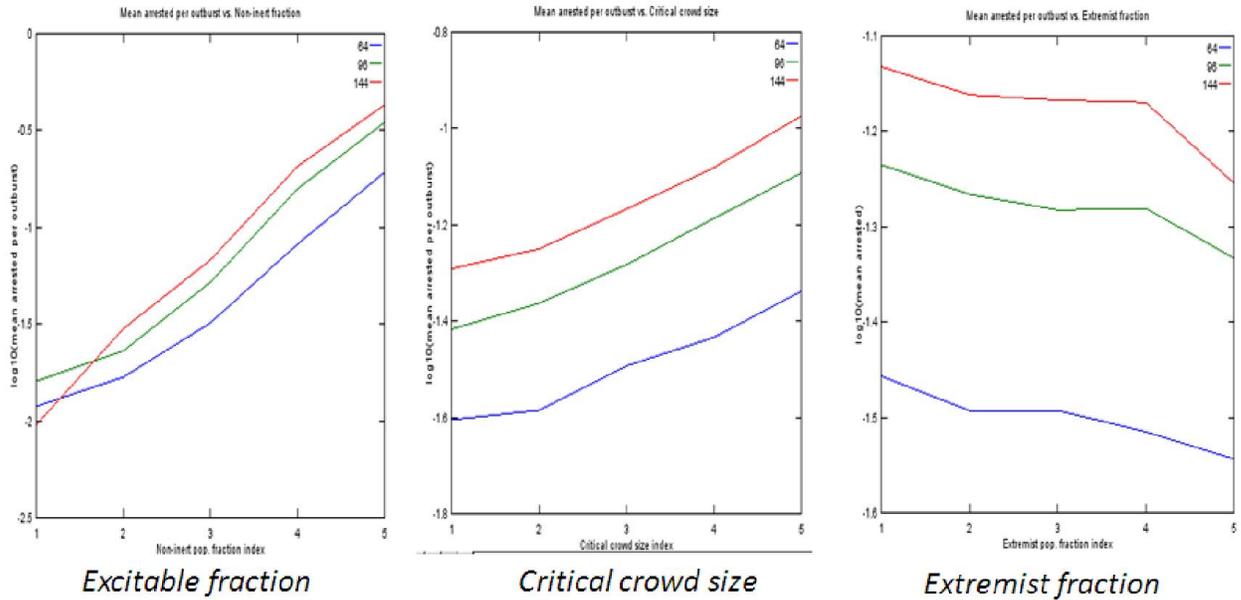

**Figure 13** Average number of arrests per protest (as a fraction of the total population) for different levels of excitable fraction (left), critical crowd size (center) and extremist fraction (right). The different curves correspond to different thresholds for minimum outburst size. The vertical scale gives the base 10 logarithm of the mean fraction of the total population arrested per protest.

The leftmost graph of Figure 13 shows a logarithmic increase in the number arrested as a function of excitable fraction, indicating an *exponential dependence* of the number arrested on the excitable population. This reinforces our findings from Figure 12 that small increases in excitable population can lead to disproportionately large increases in the severity of protests. A comparison of Figure 12 and Figure 13 suggests however that the mean protest duration is less affected than maximum protest intensity when the excitable population is increased.

The middle and rightmost graphs of Figure 13 are consistent Figure 11 and Figure 12, and our previous discussions of these two figures serve to explain the results shown here as well.

## 6 MODEL PARAMETERS: INTERPRETATION AND DETERMINATION

In this section, we consider practical issues concerning the model parameters. These include both the practical significance of the parameters, and the estimation of parameters from socioeconomic data. We conclude that at least some of the parameters are actually proxies for other effects.

### 6.1 Number of cops

As discussed in Section 5.2, outbursts in the agent-based model are entirely produced by lapses in cop coverage. These lapses occur because cop agent placement is completely random, so the number of cop agents covering any particular grid point is Poisson distributed. As a result, it is routinely the case that some grid points are covered by 8 or more agents, while others are covered by none. Such a distribution is unrealistic, if cop agents are deployed in any sort of organized manner.

This unrealistic model of cop coverage gives a misleading picture of the effect of the number of cops on the rate and severity of outbursts. For instance, we showed in Section 5.2 that a relatively small decrease in the number of cop agents in the model leads to large decrease in the mean waiting time between outbursts. Epstein cites this effect as evidence that liberalization of a society is exceptionally liable to produce violent outbursts ([1], page 2747). However this effect is a direct result of the model's unrealistic mechanism for generating outbursts. Since this mechanism is not realistic, the result is called into question.

Historically, outbursts of violence have a variety of causes which are hard to characterize. Often they are mediated by an event that produces widespread outrage: some examples are given in Table 4.

**Table 4 Riot causes for various riots**

| Riot/demonstration | Inciting event |
|---|---|
| Watts (1969) | Drunk driving arrest witnessed by bystanders |
| Mumbai riots (1992) | Destruction of Babri Masjid mosque |
| Tunisia (2010) | Immolation of Mohamed Bouazizi |
| London (2011) | Shooting death of Mark Duggan |

In summary, the "number of cops" parameter should not be interpreted literally: rather, to some extent it serves as a proxy for more complicated riot causes.

### 6.2 Jail term

Within the model, the "jail term" represents a period following arrest during which an agent is barred from activism. However, practically there are other mechanisms by which activists may be forced to be inactive for a period of time. For example, following a riot a participant may become a fugitive from justice; or he may wish to maintain a low profile. For these reasons, the model's jail term should not be interpreted literally: the effective jail term comprises a combination of effects. For this reason, direct measurement of the jail term from socioeconomic data may prove difficult.

Other properties of the model lend additional support to the assertion that the jail term cannot be interpreted literally. First, extremist agents that are released from jail immediately become active, and are immediately re-arrested. Certainly this feature of the model is not realistic. Second, an unrealistically large fraction of the activists involved in protests are arrested. In simulations, we found that the number of arrested agents in an outburst is roughly comparable with the maximum participation rate (and roughly 1/3 of the total population involved in the protest). In historical riots, the percentage of participants that are actually arrested is much smaller than this. For instance, in the 2011 London riots only about 2,000 arrests were made [7] (many of which came after-the-fact due to police use of CCTV [8]), and only about 1,300 were actually sentenced to prison [10]. On the other hand, during the riots close to 3,000 businesses and homes were attacked by multiple attackers [7]. This indicates that far less than 1/3 of the participants were physically detained, in contradiction to the agent-based model's predictions.

### 6.3 Hardship and legitimacy

Sociologists have defined various measures of hardship and legitimacy for various purposes. However, there is no empirical evidence that any of these are suitable for use as the $H$ and $L$ required by the agent-based model. In fact, the expression of grievance as a product of hardship and illegitimacy is based solely on a plausibility argument by Epstein, and has never been verified by any empirical studies. As far as risk aversion and activism threshold are concerned, these are particular to Epstein's model and have never been subject to statistical study using real data; so the determinative socioeconomic factors for $R$ and $T$ are so far completely unknown.

Finding suitable expressions for $H$ and $L$ would require cross-national statistical studies analogous to the Political Instability Task Force (PITF) study of causes of civil war (see reference [6]). Such studies would face far greater challenges than the PITF study for the following reasons:

- Riot databases are less complete and of lower quality than civil war databases, since riots are less defined, more variable, and more likely to go unreported than civil wars. Lower-quality data lead to leads to less-dependable results.
- PITF was only concerned whether or not civil wars occurred, so it obtained a *classification* of nations into "likely" and "unlikely" categories. However, the agent-based model is seeking much more detailed information, namely the *frequency* and *size distribution* of riots for each individual nation. The fact that we are trying to obtain much more detailed information implies that much better data is required to obtain this information with comparable accuracy.
- The agent-based model assigns $H$ and $L$ values to agents on an *individual* basis (as opposed to PITF which assigned likelihoods of civil war on a *national* basis). We may assume that hardship and legitimacy for individuals depends on *individual* variables (such as income) in the same way that hardship and legitimacy for nations depends on *national* variables (such as income per capita), but this is no doubt an approximation, and the accuracy of this approximation is unknown.

In view of these considerable difficulties, it is highly questionable whether statistically meaningful results can be obtained based on current data.

One possible approach would be to assume that the causative factors for riots are similar to those for civil wars, and attempt to reuse the PITF results. Unfortunately, at least one study has shown [11] that the assumption that civil wars and riots are correlated is not well-justified.

In summary, until now there is little empirical support for any specific expressions for the model inputs $H$ and $L$ in terms of measurable socio-economic quantities. Furthermore, the internal relation between the model parameters $H, L,$ and $G$ has not been empirically justified. Finally, there is no practical or theoretical grounding for determining the value of $T$.

### 6.4 Time step and grid spacing

The agent-based model locates agents on a grid and updates agents' state each time step. Since all model calculations are made relative to this grid, it follows that choice of time step and grid spacing are key determinants for model behavior and performance. As we shall see in this section, this choice also raises issues that suggest serious obstacles to practical modeling.

Since the model only updates every time step, it cannot capture behavior that occurs on smaller time scales than a single time step. So if we want to model riots, which typically initiate and develop over a period of several hours, we must choose a time step of one hour or less in order to capture the evolution of violent outbursts. On the other hand, if model time at this granularity, we must recognize that a number of other factors not currently implemented must be accounted for in order to preserve any semblance of realism. For instance, multi-day riots typically taper off in the early morning as rioters return home to sleep, only to flare up again the following night. The current agent-based model has no mechanism to account for this.

Similar issues are involved in the choice of grid spacing. Grid spacing must be somewhat less than agent vision in order for agents to be aware of other agents in their vicinity. Furthermore, if the time step is small, then agent vision is correspondingly small: probably on the order of a kilometer or less, if the time step is less than an hour. Furthermore, the grid spacing must be small enough to capture the scale of protest events. Historically, significant violent protests involving thousands of people (for example, the 2011 protests in Tahrir square, Cairo) have occurred within a one-kilometer radius – to capture these events, the grid spacing be on the order of a few meters. This would lead to impossibly large grid sizes. For instance, a single small city of 100 square kilometers modeled with a 5-meter grid spacing and 1-hour time step would involve 4 million grid points per time step, or 35 billion grid points for one year of simulation.

### 6.5 Agent motion parameters

Agent motion also raises significant problems for modeling. In the short term during actual protests, protestors may travel several kilometers to "follow the action," suggesting that agents should be mobile within the model. However, in the long term, agents will remain close to home, suggesting that agents should be stationary. These conflicting long-term and short-term motion requirements pose a serious problem to the agent-based model. It seems that an agent-based model that is suitable for simulating local, temporary protests will not be suitable for the long-term evolution of the society in which these protests take place.

## 7 CONCLUSIONS

Epstein's agent-based model does indeed exhibit characteristics that may be found in real-world examples of mass civil violence, including riots. However, it appears that many of these similarities are the result of generic mathematical properties of the model, and are not due to the fact that the model gives an accurate causal explanation of the situation. Thus the model's qualitative successes in producing riot-like phenomena in no way guarantees that the model's parameters have any realistic relationship with the parameters they nominally represent. On the contrary, we argue that the model's parameters are proxies which reflect the overall situation of the population rather than specific characteristics that can be measured by direct observation.

The mathematical characteristics of the model give insight into the performance prospects of any mathematical model of riots, agent-based or otherwise. We argue that our results indicate that riots fall within the classification of self-organized critical phenomena. One universal characteristic of such phenomena is inherent unpredictability. When modeling such phenomena, expected rates of occurrence may be estimated and high-risk situations can be identified; but high-risk situations may experience no significant events over long periods, while serious events may arise in low-risk situations with little or no warning. Thus any mathematical model of riots should be expected to be intrinsically limited in its ability to forecast. It is instructive to compare the similar field of seismology, where decades of intensive research and massive data-gathering with sophisticated instrumentation has not yielded any causal models, and regional risk assessment depends largely on the past history of regional seismic events [12].

As far as serving as a practical model of civil violence, the agent-based model of civil violence has serious structural limitations. It is only applicable to very limited forms of civil violence, where agents do not engage in any strategic planning but rather act purely on the basis of their immediate environment. Indeed, it could be argued that such is never the case. Police will inevitably have some sort of cooperative strategy, such as cordoning off protestors. Protestors will talk to each other (either by word-of-mouth or electronically) and mutually decide on targets or venues for congregation. The model also posits a geographical distribution of at most one agent per grid point, which cannot possibly reflect extreme variations in crowd density that routinely occur in riots. There are serious obstacles to choosing a space/time discretization that both captures the dynamics of riot development in the short term, and tracks population trends in the long term. The same is true for agent and cop motion rules.

The model assumptions rely on oversimplifications which severely undermine its credibility as a quantitative model. It assumes random cop distributions are far from accurate. Furthermore, it posits that the excitable population is always ready to riot, and is only inhibited by the presence of cop agents. This does not jibe with historical incidents such as the 2011 London riots or the earlier Los Angeles riots, where people decide to participate in riots because their peers are doing it and not simply because there are no cops around.

There are serious practical and so far unresolved difficulties in setting time step, grid size, and agent motion rules. Optimizing these parameters for small-scale, local demonstrations will produce a model that is inappropriate for long-term evolution, and vice versa.

Some accommodations can be made for the model's current limitations. The behavior rules can be modified so that long-distance influences are taken into account (see [13]). It is conceivable that strategy could be introduced into the behavior rules for cop and populace agents, but so far this prospect remains uninvestigated. The notion of grievance can be improved to more accurately reflect observed fluctuations based on current events. Rather than using socioeconomic data to determine model parameters, past history of riots may be used to determine expected riot frequency and size, and a calibration model could be used to adjust model parameters that match the expected behavior. However, the serious intrinsic weaknesses of the model calls into question the model's prospects for quantitative accuracy. It appears the model is more suitable for a heuristic understanding of qualitative behavior than for a quantitative characterization of real-life systems.

**ACKNOWLEDGMENTS**

This work was partially supported by the Air Force Research Laboratory (AFRL) under a grant from the Air Force Summer Faculty Fellowship Program (SFFP). This paper has been approved for public release (Case #88ABW-2013-0828).

The investigators greatly benefitted from collaboration and discussions with the following individuals: John Salerno, Jason Smith, and Adam Kwiat.**REFERENCES**